\documentclass[
    ,final            
  ]
  {aipproc}

\layoutstyle{6x9}

\begin{document}

\title{The Atacama Cosmology Telescope}

\author{Arthur Kosowsky}{
   address={Department of Physics and Astronomy, Rutgers University, 136 Frelinghuysen Road, Piscataway NJ 08854-8019}
}





\begin{abstract}
The Atacama Cosmology Telescope (ACT) project is described.
This multi-institution collaboration aims to produce
arcminute-resolution and micro-Kelvin sensitivity maps of the
microwave background temperature over 200 square degrees of the sky in
three frequency bands. We give a brief overview of the scientific
motivations for such a map, followed by a design outline of our
six-meter custom telescope, an overview of our proposed bolometer
array detector technology, and site considerations and scan strategy.
We also describe associated optical and X-ray galaxy cluster
surveys.
\end{abstract}

\maketitle

\section{Scientific Motivation}

With results from WMAP in hand, it is clear that the near-term future
of microwave background measurements will be primarily a push towards
smaller angular scales and polarization.  As described at this
meeting, many small-scale experiments are currently underway, under
construction, or in the planning stage. This paper describes an
ambitious proposed collaboration, the Atacama Cosmology Telescope,
which aims to combine new bolometer array technology with a
custom-designed six-meter telescope to produce an instrument with a
notable combination of sensitivity, angular resolution, and
control of systematic errors.  Current information about the
experiment is available at
http://www.hep.upenn.edu/~angelica/act/act.html.  Before detailing the
experimental and observational aspects of this effort, we give a brief
summary of the main scientific questions we aim to address, which are
covered in more detail elsewhere in these proceedings (see also
\cite{kos03}).

At angular scales smaller than around 4 arcminutes, corresponding to
multipoles $l>3000$, nonlinear contributions begin to dominate the
total microwave background temperature anisotropies. The major sources
of temperature fluctuations on these scales include the
Sunyaev-Zeldovich effect, the Ostriker-Vishniac effect, and
gravitational lensing. All of these effects arise both from individual
clusters of galaxies and from the large-scale matter distribution.

The largest amplitude signal will come from the thermal SZ galaxy
cluster distortions. We expect to compile a large catalog of clusters
selected by their SZ signals, which provides a cleaner cluster
selection criterion than flux-limited optical or X-ray surveys. This
catalog will be well-suited to measuring the cluster number density as a
function of mass and redshift, which is a sensitive probe of the
growth rate of structure since redshift $z=1$; in turn, the
structure growth rate constrains dark
energy and neutrino masses. We also expect to place significant
constraints on cluster masses and peculiar velocities via their
kinematic SZ and gravitational lensing signatures. For the diffuse
signals, we aim to detect gravitational lensing of the microwave
background and construct a projected mass map on scales of 15
arcminutes. We expect to have sufficient sensitivity to detect the
Ostriker-Vishniac effect, which is sensitive to the redshift and
spatial variation of reionization, and the Rees-Sciama effect, which
is sensitive to the non-linear evolution of gravitational potentials.

ACT will provide a measurement of the power spectrum on all scales
from $l=200$ to $l=10000$, probing the primordial fluctuation spectrum
for departures from a power law or for features; either could arise
from non-minimal models of inflation. Having a single experiment which
probes a wide range of angular scales with good control of systematics
is crucial for uncovering small departures from perfect power law
primordial spectra.

ACT's scan strategy results in a significant area of the survey being
in the galactic plane. A variety of interesting topics in galactic
astrophysics can be addressed with such a map, particularly properties
and distribution of dust.

\section{Telescope}

The stringent control of systematic errors necessary to attain
temperature measurements at 1 $\mu$K sensitivity drives many aspects
of our telescope design. A schematic illustrating important features
is shown in Fig.~1. The primary reflector has a diameter of 6 meters,
producing diffraction-limited resolution of approximately 1.7
arcminutes at 150 GHz and 0.9 arcminutes at 270 GHz. A preliminary
design configuration gave a clear aperture by
employing three off-axis, aspherical mirrors in a double-Gregorian
configuration; the final optical design will be somewhat different.
The resulting field of view is wide and flat, with
minimal side-lobe response and telescope offset.  Additional reimaging
optics and a cold Lyot stop will be contained within the cryostat
containing the detectors described below.

\begin{figure}
\includegraphics[height=.4\textheight]{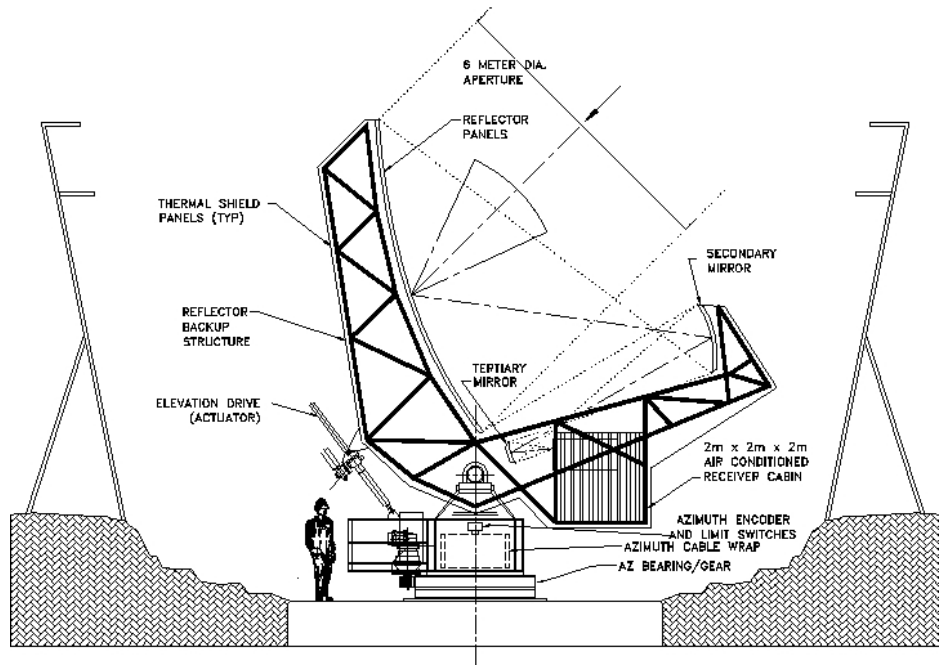}
\caption{A schematic design for the Atacama Cosmology Telescope.
For clarity of presentation, an additional ground screen which rotates
in azimuth with the telescope is not shown. The final optical design
of the telescope will be different.}
\end{figure}

The telescope is low to the ground, allowing large shields to
reduce ground pickup.  The entire optical path will maintain
polarization information; while our initial plans do not include
polarized detectors, our design will allow future detector upgrades.

The telescope sits on a large bearing; the entire telescope structure,
including a ground screen not shown in the figure, will rotate in
azimuth. This strategy guarantees that no reflecting or diffracting
surfaces ever move with respect to the receiver, and thus any pickup
through the near-field sidelobes
will remain constant. The nominal observing elevation of the
telescope is $45^\circ$ with limited elevation control,
but as explained below, the baseline
observing strategy calls for a constant elevation so the same
atmospheric depth is observed at all times.

The telescope will be remote controlled with data transferred via a
radio downlink, since the Atacama site (described below) is at a
high altitude. Such a scheme
has been previously employed with the MAT/TOCO experiment \cite{mil02}
by members of our collaboration.

\section{Detectors}

Current experiments employ arrays as large as roughly a hundred
bolometers.  The ``camera'' we propose as the receiver
for the ACT telescope will employ three separate $32\times 32$ arrays
of 1 square-millimeter superconducting transition edge sensor (TES)
bolometers \cite{irw96,lee96} on pop-up arrays . In recent years,
Harvey Moseley and colleagues at Goddard Space Flight Center have
developed a technology in which bolometers are
packed into CCD-like arrays. Arrays have so far been fabricated for
use at shorter wavelengths; see \cite{dow02} for an overview of the
$12\times 32$ SHARC array which has been built for the Caltech
Submillimeter Observatory. These detector arrays have a number of
distinct advantages: (1) The array elements are close-packed,
filling the focal plane; (2) The TES
thermal time constant is $\leq 1.5$ ms ($f_{\rm max} \approx 110$ Hz),
allowing rapid scanning of the arrays; (3) Each TES is intrinsically
stable on time scales of 0.8 s ($f_{\rm knee} \approx 0.2$ Hz); (4)
The TES intrinsic electrical noise equivalent power is below
$10^{-17}$ ${\rm W}/\sqrt{\rm Hz}$ at 265 mK physical temperature.
The detector arrays will be maintained at temperatures near 270 mK
using a combination of pumped $^4$He and
$^3$He systems, with no consumable cryogens.  To minimize the number
of connections to the dewar, we plan to read out each array with a
SQUID multiplexor \cite{che99} which has been developed by Kent
Irwin's group at NIST.

We plan to observe three frequency bands simultaneously: a low
frequency band centered around 145 GHz, a middle band around 225 GHz,
and a high frequency band around 265 GHz; the width of each band will
be 25 to 30 GHz. The middle frequency band will straddle the null of
the thermal SZ effect, allowing a clear identification of SZ
galaxy clusters via their signal in all three frequency bands. Our
target sensitivity per bolometer is 300 $\mu$K sec$^{1/2}$ in the
lowest frequency band, 500 $\mu$K sec$^{1/2}$ in the middle band, and
700 $\mu$K sec$^{1/2}$ in the high frequency band.

\section{Site Characteristics}

The telescope will be sited on Cerro Toco, in the Atacama desert of
the Chilean Andes, at an altitude of 5200 meters. The site is the same
one used previously for the MAT/TOCO experiment \cite{mil02}, and is a
few kilometers from the ALMA \cite{woo02} and CBI \cite{pad02} site. The
weather and atmosphere at these sites is well characterized;
atmospheric measurements over several years at a frequency of 225 GHz
are available at the ALMA web site, http://www.alma.nrao.edu. The
MAT/TOCO experiment also operated at the ACT site during 1997 and 1998
and made extensive atmospheric measurements; the median measured sky
temperature during these observing runs was approximately 9 K at 150 GHz.

The atmospheric characteristics of our site are comparable to those of
the South Pole, where several other small-scale microwave background
measurements will be sited. The ACT site is at a significantly higher
altitute, but usually has a higher amount of precipitable water vapor.
While the South Pole has a lower yearly-averaged atmospheric opacity
than the ACT site, well-understood annual and diurnal fluctuations at
the Chile site bias this comparison. For roughly 8 months per year,
the opacities at the two sites are comparable, and when night-only
data is considered, the opacity in Chile is actually better than that
at the South Pole 25\% of the time. ACT will observe only during the 6
months of each year in which atmospheric conditions are most
favorable.

We have used the calculations of Lay and Halverson \cite{lay00} for
the expected noise contribution from atmospheric emission to a scanned
beam to estimate the atmosphere-induced temperature fluctuations for
ACT as a function of angular scale. We find that for angular scales
smaller than about 25 arcminutes, the atmospheric noise will be
smaller than our receiver noise. For this estimate, we have used the
best 50\% of all the atmospheric data.

\section{Scan Pattern and Projected Sensitivity}

The major advantage of the ACT site is geographic. A change in
observing elevation by 1 degree can result in a change in the
effective atmospheric emission of tens of milli-Kelvin due to the
changing atmospheric column density, which is a factor of $10^4$
larger than the fluctuations we are trying to measure. To control
systematic errors and striping in the maps, observations at constant
elevation are highly beneficial. While a scan strategy has not been
finalized, we anticipate something close to the following.  The site
is at $23^\circ$ south latitute; our first observations are planned at
the nominal fixed elevation of $45^\circ$. As an example strategy, we
scan the entire telescope $\pm 1.5^\circ$ in azimuth centered on one
of two azimuths at $207^\circ$ and $153^\circ$.  This
constant-elevation scan pattern over 24 hours results in a
cross-linked pattern over an annulus $1.7^\circ$ wide and $140^\circ$
long.  The cross-linking facilitates a robust map solution and allows
for systematic checks.  Approximately half of the survey region will
be through the galactic plane, while the other half, around 100 square
degrees, will be suitable for cosmological analysis.  Primary
calibration will be via comparison with the WMAP maps; we will also
have the capability to track in elevation for observation of secondary
calibration sources.

The combination of optical geometry, detector configuration, and scan
strategy results in each sky pixel being observed on 6 distinct time
scales: (1) Each detector pixel scans across a sky pixel in 0.02
seconds; (2) The entire detector array scans across a sky pixel in 0.4
seconds; (3) Each complete scan of the telescope is completed in 3
seconds; (4) A given sky pixel drifts through the size of the beam in
9 seconds; (5) A given sky pixel drifts throught the size of the field
of view in 3 minutes; (6) A given pixel is observed at an interval of
7 hours as it rotates from one azimuthal chop region to the
other. This chopping scheme is made possible by the fast response,
sensitivity, and stability of the TES bolometers. Using the strong
spatio-temporal filter provided by the scan strategy, we aim to solve
for the celestial signal in the presence of atmospheric and
instrumental fluctuations. Recall that the scan strategy is
accomplished by moving the entire telescope, with the optical path
kept fixed at all times.

We plan science observations only during half of the year corresponding
to the most favorable atmospheric conditions.
With the combination of telescope, detectors, and observation strategy
described here, our target sensitivity can produce a map with 2 $\mu$K
errors per $1.7'$ pixel over 200 square degrees.

\section{Associated Galaxy Cluster Surveys}

ACT will identify on the order of a thousand galaxy clusters in the
survey region (depending on the cosmological model), down to a
limiting mass of around $2\times 10^{14} M_\odot$ to arbitrary
redshift.  Use of this cluster catalog as a cosmological probe
requires cluster redshifts. We plan to use the Prime Focus Imaging
Spectrograph on the 11-meter Southern African Large Telescope (see
http://www.salt.ac.za), currently under construction at the South
African Astronomical Observatory, to obtain up to 400 cluster
redshifts, along with galaxy velocity dispersions, over two years. By
using a slit mask, SALT will be able to determine spectroscopic
redshifts for 30 cluster galaxies simultaneously in 30 minute
exposures for clusters at redshifts $z < 1$; the galaxy velocity
dispersion will be determined along with the mean redshift, providing
an independent estimate of cluster mass.  Rutgers, which is a 10\%
partner in building SALT, has committed 10 nights of observing time
per year to cluster follow-up observations. Additional spectroscopic
redshifts as well as photometry will be obtained by our Chilean
collaborators using various large telescopes.

We also plan to pursue X-ray imaging of a subset of clusters with the
Chandra and XMM satellites, to obtain information on cluster gas
temperatures and densities. X-ray measurements will provided estimates
of cluster masses, densities, and temperatures, although subject to
some systematic errors due to cluster substructure and departures from
isothermality.  Information from X-ray observations will help to
establish the connection between cluster SZ signals and cluster
masses. High-resolution optical imaging with the Hubble Space
Telescope would also provide a wealth of data for probing evolution of
galaxies in clusters at a wide range of redshifts, and we would like to
initiate such an observing program, if possible,
once a reliable cluster catalog has been established.

\section{Current Status and Timeline}

ACT has been funded by the NSF beginning in January 2004.  
Much development work has already occurred, particularly in the
areas of optical design, cryogenic design, and bolometer array testing.
We anticipate that telescope construction will take two years and
receiver design and fabrication 3 years, followed by two full seasons
of observing. The SALT telescope and prime focus spectrograph are on
budget and on schedule for first light in February 2005.

The next generation of microwave background measurements at
unprecedented sensitivities are now on the horizon, making use of new
technologies and past design experience.  We anticipate that the
Atacama Cosmology Telescope project described here will produce a
high-fidelity map of the cosmic microwave background radiation at
arcminute resolution and near micro-Kelvin sensitivities, opening the
door to a number of new and interesting cosmological probes.

\begin{theacknowledgments}

The ACT governing board is composed of Mark Devlin, Kent Irwin,
Arthur Kosowsky, Harvey Moseley,
Lymna Page (PI), David Spergel, and Suzanne Staggs. Initial planning
was supported by NSF award PHY00-99493 to Lyman Page at Princeton and
NSF award AST97-32960 to Mark Devlin at the University of Pennsylvania
along with the University of Pennsylvania Research
Foundation. Additional collaborators are at Princeton University, University
of Pennsylvania, Rutgers University,
NASA Goddard Space Flight Center, National Institute of Standards and
Technology, Haverford College, University of
Toronto, Columbia University, Drexel University, Catolica University
(Chile), and University of Cardiff (UK). Lyman Page and Mark
Devlin provided helpful critiques of an earlier draft of this paper.
A.K. is a Cottrell Scholar of the Research Corporation.
\end{theacknowledgments}

\end{document}